\documentclass[fleqn,12pt,twoside]{article}
\usepackage{espcrc1}

\usepackage{graphicx}
\usepackage{epsf,epsfig,psfig,float,amssymb,stmaryrd,latexsym}
\usepackage{graphics,psfrag}
\usepackage[figuresright]{rotating}

\def\3{{\ss}}

\def\vek #1 {\overrightarrow {#1}}

\newcommand{\beq}{\begin{equation}}
\newcommand{\eeq}{\end{equation}}
\newcommand{\beqn}{\begin{displaymath}}
\newcommand{\eeqn}{\end{displaymath}}
\newcommand{\beqa}{\begin{eqnarray}}
\newcommand{\eeqa}{\end{eqnarray}}
\newcommand{\beqan}{\begin{eqnarray*}}
\newcommand{\eeqan}{\end{eqnarray*}}

\newcommand{\bma}{\begin{array}{cc}}
\newcommand{\ema}{\end{array}}

\newcommand{\AmS}{{\protect\the\textfont2
  A\kern-.1667em\lower.5ex\hbox{M}\kern-.125emS}}

\title{Chiral effective field theory for few--nucleon systems}

\author{E. Epelbaum\address{Ruhr-Universit\"at Bochum, Institut f{\"u}r
  Theoretische Physik II, \\ D-44870 Bochum, Germany}}

\begin{document}

\maketitle

\begin{abstract}
Some recent developments in the description of nuclear forces and 
few--nucleon systems within the effective field theory approach are reviewed. 
\end{abstract}

\section{INTRODUCTION}
Chiral Effective Field Theory (EFT) has become a standard tool for analyzing the 
properties of hadronic systems at low energy where the perturbative 
expansion of Quantum Chromodynamics (QCD) in powers of the coupling 
constant cannot be applied, see \cite{Ulf,Silas}.  It is based upon the approximate and 
spontaneously broken chiral symmetry of QCD. Starting from the most 
general effective Lagrangian for Goldstone bosons (pions in the two--flavor
case of $u$ and $d$ quarks) and matter fields (nucleons, $\Delta$, $\ldots$)
consistent with the symmetries of QCD,
the hadronic S--matrix elements are obtained via a simultaneous expansion
in the low external momenta and quark masses. Goldstone boson loops are naturally 
incorporated and all corresponding ultraviolet divergences can be absorbed 
at each fixed order in the chiral expansion by the counter terms of the 
effective Lagrangian.

This perturbative scheme works well in the pion--nucleon and pion--pion sectors, 
where the interaction vanishes at vanishing external momenta in the chiral limit.
In the case of few interacting nucleons one 
has to deal with non--perturbative problems. Indeed, perturbation theory is 
expected to fail already at low energy due to the presence of the shallow 
few--nucleon bound states. A suitable non--perturbative approach has been 
suggested by Weinberg \cite{Weinb}, who showed that the strong 
enhancement of the few--nucleon 
scattering amplitude arises from purely nucleonic intermediate states. 
Weinberg suggested to apply EFT to the kernel of the corresponding scattering   
equation, which can be viewed as an effective nuclear potential. 
This idea has been explored in the last decade by many authors.
In the following I will briefly address some of the actual topics related to 
the EFT description of few--nucleon systems.

\section{FEW NUCLEONS AT NEXT--TO--NEXT--TO--LEADING ORDER}

The procedure suggested by Weinberg
has been first carried out for two nucleons by  Ord\'o\~nez and co--workers, 
who derived a NN potential up to next--to--next--to--leading 
order (NNLO) in the chiral expansion and performed a 
numerical analysis of the two--nucleon (2N) system \cite{Ordonez96}.
The explicit energy dependence of the effective potential derived 
in \cite{Ordonez96} leads, however, to difficulties 
in applications to $>2 N$ systems but can be eliminated by certain techniques. 
The corresponding energy--independent expressions for 
the NN potential have been first given 
in ref.~\cite{friar94} and later in refs.~\cite{kaiser97,egm98} using different 
methods. They served as a basis for the NNLO analyses of the two--nucleon (2N) system
\cite{egm00,ep02}, see also \cite{ent01} for related application, 
as well as three-- (3N) and four--nucleon (4N) systems \cite{ep02a} 
including the corresponding 3N forces. 
\begin{figure}[tb]
\centerline{
\psfig{file=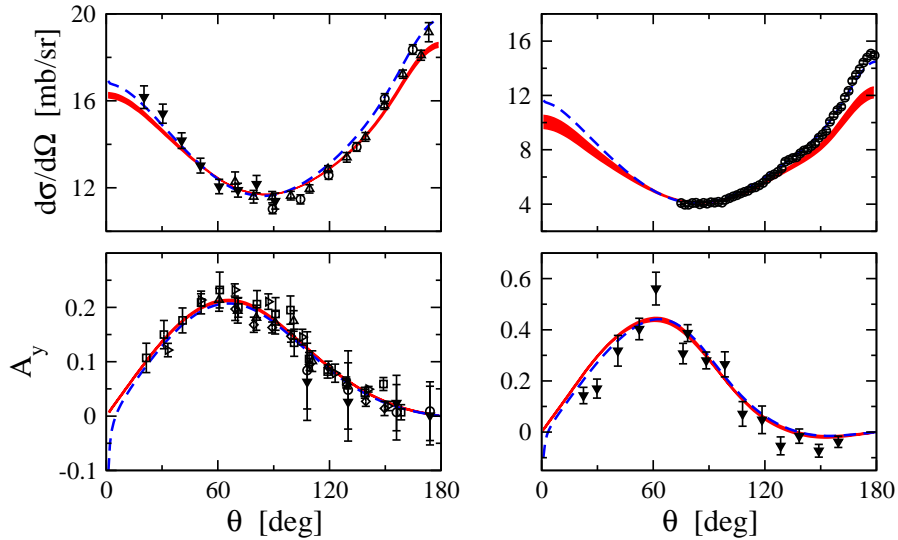,width=12.cm}}
\vskip -0.7 true cm
\centerline{
\parbox{1.00\textwidth}{
\caption[fig1]{
\label{fig1}  
{\it np} differential cross section (upper row) and vector analyzing power 
(lower row) at 
$E_{\rm lab}=50$ MeV (left panel) and $E_{\rm lab}=96$ MeV (right panel).
Shaded band refers to the NNLO result, dashed line to the Nijmegen phase shift analysis
(NPSA) \cite{npsa}. References to data can be found in \cite{npsa}.
}}}
\end{figure}

According to chiral power counting the dominant contributions
to the effective Hamiltonian for few nucleons are of the order $Q^0/\Lambda^0$, 
where $Q\sim M_\pi$ refers to the soft scale (typical momenta involved in the 
process) and $\Lambda$ to the hard scale (chiral symmetry breaking scale, 
ultraviolet cut--off)  \cite{Weinb}. These leading--order (LO) contributions turn out to be 
of  the 2N type, given by one--pion exchange (OPE) and two NN contact interactions   
without derivatives. The first corrections at next--to--leading order (NLO) 
are of the order $Q^2/\Lambda^2$ and still of the 2N type. 
They result from two--pion exchange (TPE) with the leading  (i.e. with one derivative) 
$\pi NN$ and $\pi \pi NN$ vertices and 7 independent
NN contact interactions with two derivatives.\footnote{The additional two 
momentum--independent contact interactions with one insertion of $M_\pi^2$
can be absorbed by the LO contact operators.} The only low--energy constants 
(LECs) entering the expressions for TPE at NLO are the nucleon axial--vector  
coupling $g_A$ and the pion decay constant $F_\pi$. Both LECs are measured rather 
accurately, so that the leading TPE contribution is parameter--free. On the contrary,
the LECs accompanying the contact operators are unknown and have been fixed
from a fit to low--angular--momentum partial waves.  

At NNLO ($\sim Q^3/\Lambda^3$) one has to take into account the subleading 
TPE contributions given by the triangle diagram with the $\pi \pi NN$ vertex 
with two derivatives or one insertion of $M_\pi^2$. The corresponding LECs
are denoted $c_{1,3,4}$ and have been fixed in the $\pi N$ system, see e.g. 
\cite{paul00}. The numerical values of these LECs found in \cite{paul00} as well as 
in several other analyses of  $\pi N$ scattering are rather large compared 
to what is expected on dimensional reasons, see ref.~\cite{ep02}. 
Similar large values of $c_{3,4}$ have also been obtained recently from 
the $np$ and $pp$ partial wave analysis carried out by the Nijmegen group 
\cite{PSA_c}. The large values of the $c_{3,4}$ can at least be partially explained 
by the fact that these LECs are saturated by the $\Delta$--excitation
\cite{bern97}. This implies that a new and relatively small scale, namely $m_\Delta 
- m_N \sim 293$ MeV enters the values of these constants in EFT without explicit 
$\Delta$. The large numerical values of the $c_i$'s lead 
to the subleading TPE contribution to the NN potential 
which shows an unphysically strong attraction already at intermediate distances 
$r \sim 1 -2$ fm when standard regularization techniques to pion loop 
integrals (i.e. dimensional or infinite momentum cut--off regularization)
are applied. 
The unphysical behavior of the potential shows up most notably at NNLO in 
parameter--free predictions for D--waves (F--waves) which 
start to strongly deviate from the 
data already at $E_{\rm lab} > 50$ MeV ($E_{\rm lab} > 150$ MeV) after  
the subleading TPE contribution is taken into account \cite{kaiser97}. 
Higher order counter terms are needed in order to reduce the strong cut--off 
dependence of the D--wave phase shifts and to correct the F--waves, which 
indicates the slow convergence of the chiral expansion. In addition, 
unphysical deeply bound states arise in low partial waves \cite{egm00}. 
Although such deeply bound states do not influence NN observables at 
low energy, they might lead to certain complications in another processes like 
e.g.~$Nd$ \cite{ep02} and $\pi d$ \cite{beane03} scattering. 

In order to avoid the above mentioned difficulties we have constructed 
in ref.~\cite{ep02} the NNLO* version of the potential without deeply bound 
states. To achieve that, we adopted values of the LECs $c_{3,4}$ much smaller 
in magnitude than the ones obtained from $\pi N$ scattering.  The resulting 
potential leads to a good description of the low--energy NN data as 
exemplified with selected observables in Fig.~\ref{fig1}. 
\begin{figure}[tb]
\centerline{
\psfig{file=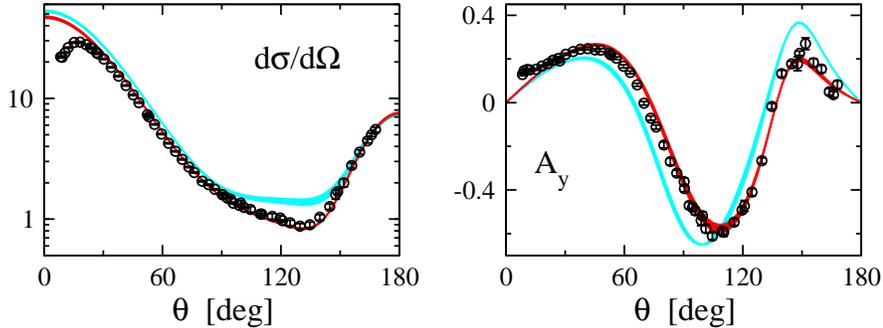,width=12.cm}}
\vskip -0.7 true cm
\centerline{
\parbox{1.00\textwidth}{
\caption[fig2]{
\label{fig2}  
{\it Nd} differential cross section and vector analyzing power at 
$E_{\rm lab}=65$ MeV.
Light (dark) shaded bands refer to the NLO (NNLO) results.
Data are from \cite{shimizu,witala}.
}}}
\end{figure}

Lot of progress has also been achieved for 3N and 4N systems. 
Scattering and bound state problems in such systems can be 
solved in a numerically exact way using Faddeev--Yakubovsky theory and
serve as a good testing ground for chiral 
forces, since most of the unknown parameters are already fixed 
in the two nucleon system.  
As already stated before, no 3N forces appear at LO and NLO. 
Parameter--free calculations for $Nd$ scattering 
as well as 3N and 4N bound states at that order in the chiral expansion
have been presented in ref.~\cite{epPRL}. We found a good description 
of various scattering observables at low energy, see Fig.\ref{fig2}.
The predicted $^3$H and $^4$He binding energies 
\beqan
\mbox{ \hskip 1 true cm BE} (^3\mbox{H}) &=& 
            -7.53 \ldots -8.54 \mbox{ MeV \hskip 1.5 true cm for }
            \Lambda = 500 \ldots 560 \mbox{ MeV}, \\
\mbox{ \hskip 1 true cm BE}(^4\mbox{He}) &=& 
            -23.87 \ldots -29.57 \mbox{ MeV \hskip 1.07 true cm for }
            \Lambda = 500 \ldots 560 \mbox{ MeV},
\eeqan
where  $\Lambda$ refers to the momentum cut--off in the 2N force,
are in a similar range 
as the ones obtained using modern phenomenological NN potentials and 
have to be compared with the empirical values 
$\mbox{BE} (^3\mbox{H}) = -8.68$ MeV and $\mbox{BE} (^4\mbox{He}) = -29.8$ 
MeV.\footnote{These values have been corrected to adjust for $pp$ and $nn$
forces missing in the calculations.} 

Chiral 3N forces start to contribute at NNLO and are 
given by TPE, OPE with the pion emitted (or absorbed)
by the NN contact interaction and 3N contact interaction. 
The TPE contribution is parameter--free. 
The Pauli principle together with the usual symmetry requirements (parity 
invariance, rotational invariance, $\ldots$) lead to a strong 
reduction of the number of independent terms in the remaining 
part of the 3N force,
leaving just one OPE and one contact operator \cite{ep02a}. 
The two corresponding unknown parameters have been fixed from 
the triton binding energy and $nd$ scattering length, which allowed us 
to make parameter--free predictions for various other observables.
For the $\alpha$--particle binding energy we find at NNLO 
\beqn
\mbox{ \hskip 1 true cm BE}(^4\mbox{He}) =
            -29.51 \ldots -29.98 \mbox{ MeV \hskip 1.07 true cm for }
            \Lambda = 500 \ldots 600 \mbox{ MeV},
\eeqn
which is rather close to the empirical value. Notice that the 
cut--off dependence is strongly reduced compared to the 
NLO result. One also observes improvement for various 3N scattering 
observables when going from NLO to NNLO, see  Fig.~\ref{fig2} for two examples.

\section{IMPROVING THE CONVERGENCE OF THE CHIRAL EXPANSION FOR NUCLEAR FORCES}

As already pointed out in the previous section, the numerically large values of the 
LECs $c_i$ found in the $\pi N$ system lead to an unphysically strong attraction 
of subleading TPE. Kaiser et al.~have performed parameter--free 
perturbative calculations of the 
peripheral NN partial waves \cite{kaiser97} 
and demonstrated that the nice agreement with the data 
in D-- and F--waves at NLO is destroyed after  
accounting for the NNLO TPE contribution, which 
might indicate problems with the convergence of the chiral expansion. 
Although these problems have been avoided in the previously discussed 
analysis of the 2N, 3N and 4N
systems by a reduction of the values of $c_{3,4}$, the situation is clearly 
not satisfactory, since the reduced values of these LECs are not compatible   
with the $\pi N$ system. 
\begin{figure}[tb]
\centerline{
\psfig{file=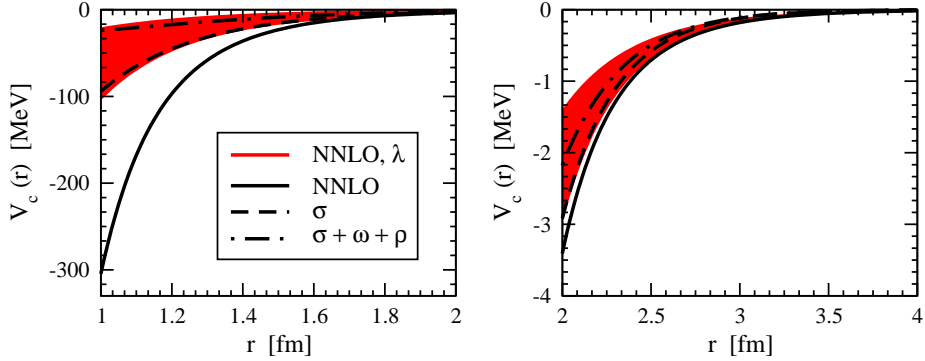,width=12.5cm}}
\vskip -0.7 true cm
\centerline{
\parbox{1.00\textwidth}{
\caption[fig3]{
\label{fig3}  The potential $V_C$  in $r$--space. 
The solid line (shaded band) shows the DR (spectral function regularized, 
$\lambda = 500 \ldots 800$ MeV) result. The dashed (dashed--dotted) line 
refers to the phenomenological $\sigma$ ($\sigma + \omega + \rho$) contributions based on the
isospin triplet configuration space version (OBEPR) of the Bonn potential \cite{Mach_rep}.   
}}}
\end{figure}

In the following I will explain the origin of the slow convergence 
and present a new method to improve it. To be specific, consider the 
isoscalar central part of the subleading TPE which results from the 
triangle diagram and is given by
\beq
\label{pot1}
V_{\rm C} (q) = \frac{3 g_A^2}{16 F_\pi^4} \int \, \frac{d^3 l}{(2 \pi)^3} 
\frac{l^2 - q^2}{((\vec q - \vec l \, )^2+ 4 M_\pi^2)((\vec q + \vec l \,)^2+ 4 M_\pi^2)} 
\left( 8 c_1 M_\pi^2 + c_3 (l^2 - q^2) \right)\,,
\eeq
where $\vec q$ is the nucleon momentum transfer and $q \equiv | \vec q \,|$, 
$l \equiv | \vec l \, |$.
The integral is cubically divergent and needs to be regularized. Applying 
dimensional regularization (DR) one finds:
\beq
\label{pot2}
V_{\rm C}  (q) = - \frac{3 g_A^2}{16 \pi F_\pi^4}
\left( 2 M_\pi^2 ( 2 c_1 - c_3 ) - c_3 q^2 \right) (2 M_\pi^2 + q^2) \frac{1}{2 q}
\arctan \frac{q}{2 M_\pi} + \ldots \;.
\eeq
The ellipses refer to polynomial (in $q^2$) terms of the kind $\alpha + \beta q^2$.
In this section I will be interested in D-- and higher partial waves, where 
such terms do not contribute. 
In order to obtain the potential in coordinate space one has to make
an inverse Fourier--transform of $V_{\rm C}  (q)$ in eq.~(\ref{pot2}). 
The ordinary  inverse Fourier--transform is obviously not possible 
due to the the fact that $V_{\rm C}  (q)$
growth with $q$. One can nevertheless obtain $V_{\rm C}  (r)$
at each $r > 0$ using the spectral function representation \cite{kaiser97}:
\beq
\label{spectrfun}
V_{\rm C}  (q) = \frac{2 q^4}{\pi} \int_{2 M_\pi}^\infty d \mu \, \frac{1}{\mu^3}
\, \frac{\rho (\mu)}{\mu^2 + q^2},
\eeq
where the spectral function $\rho (\mu )$ can be obtained from 
$V_{\rm C} (q )$ in eq.~(\ref{pot2}) via
\beq
\label{rho}
\rho (\mu ) = \Im \left[ V_{\rm C} (0^+ - i \mu ) \right]
= - \frac{3 g_A^2}{64 F_\pi^4} \left( 2 M_\pi^2 ( 2 c_1 - c_3) + c_3 \mu^2 \right)
(2 M_\pi^2- \mu^2) \frac{1}{\mu} \theta (\mu - 2 M_\pi )\,.
\eeq
In eq.~(\ref{spectrfun}) the twice subtracted dispersion integral
is given which is needed in order to account for the large--$\mu$
behavior of $\rho (\mu)$.
\begin{figure}[tb]
\centerline{
\psfig{file=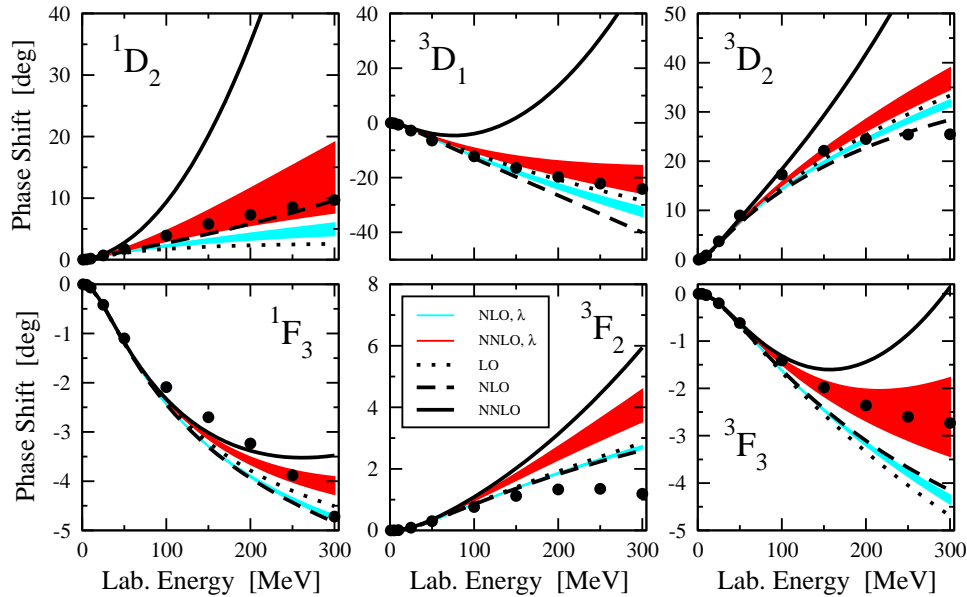,width=13.0cm}}
\vskip -0.7 true cm
\centerline{
\parbox{1.00\textwidth}{
\caption[fig4]{
\label{fig4}  
Peripheral NN phase shifts. 
Dotted line refer to LO, dashed and solid 
lines (light and dark shaded bands) to NLO and NNLO calculations using 
DR (the new regularization scheme). Solid dotes are the results from 
Nijmegen phase shift analysis.
}}}
\end{figure}

The inverse Fourier--transform in terms of the spectral function 
$\rho (\mu )$ can easily be evaluated via
\beq
\label{four}
V_{\rm C} (r ) = \frac{1}{2 \pi^2 r} \int_{2 M_\pi}^\infty d \mu
\, \mu \, e^{- \mu r} \rho (\mu ).
\eeq
Substituting $\rho (\mu )$ from eq.~(\ref{rho}) into eq.~(\ref{four})
and using for the LECs $c_{1,3}$ the central values from ref.~\cite{paul00},
$c_1 = -0.81$ GeV$^{-1}$ and $c_3 = -4.70$ GeV$^{-1}$,
one obtains the coordinate space representation of the potential
$V_{\rm C} (r )$ shown by the solid line 
in Fig.~\ref{fig3}. The central part of the NNLO TPE
potential turns out to be several times more attractive 
at intermediate distances than the 
phenomenological $\sigma$ ($\sigma + \omega + \rho$) contributions. 
This unphysical attraction shows up in the D-- and F--wave phase shifts 
as depicted in Fig.~\ref{fig4}, which have been calculated using the Born 
approximation. While phase shifts at LO (based on OPE) and NLO (based on 
OPE $+$ leading TPE) are in a fair agreement with the data, NNLO results 
(based on OPE $+$ leading TPE $+$ subleading TPE)
disagree  significantly from the data.

The origin of the unphysical attraction at NNLO 
can be traced back by looking at the integral in eq.~(\ref{four}).  
In Fig. 5 the integrand in eq.~(\ref{four}) is plotted 
versus $\mu$ at large ($r = 2 M_\pi^{-1}$), intermediate  
($r = M_\pi^{-1}$) and short  ($r = 0.5 M_\pi^{-1}$) distances.
While at large distances the integral is dominated by  low--$\mu$
(of the order $\mu \sim 350$ MeV) components, already at intermediate 
distances  rather high--$\mu$ (of the order  $\mu \sim 600$ MeV) 
components in the spectral function provide a dominant contribution. 
Clearly, at shorter distances even higher--$\mu$ components become 
important. Chiral EFT can hardly provide convergent results 
for the spectral function at $\mu \sim 600$ MeV and higher. Instead 
of keeping such large--$\mu$ components in the regularized loop integral 
expressions it is advantageous to perform the integration in eq.~(\ref{four}) 
only over the low--$\mu$ region, where chiral EFT is applicable. 
This can be achieved by introducing the regularized spectral function 
\beq
\label{regspectr}
\rho (\mu ) \rightarrow \rho^\lambda (\mu ) = \rho (\mu ) \, \theta (\lambda - \mu )\,,
\eeq
with the reasonably chosen cut--off $\lambda < M_\rho$. 
Certainly, taking a too small $\lambda$ in eq.~(\ref{regspectr}) will remove 
the truly long--distance physics while too large values for the cut--off 
may affect the convergence of the EFT expansion due to inclusion of spurious 
short--distance physics. Notice that a very similar idea with the finite 
momentum cut--off has been successfully applied to improve the convergence
of the SU(3) baryon chiral perturbation theory \cite{don98,bor02}.

In Fig.~\ref{fig3} we show $V_{\rm C} (r)$ obtained 
using the spectral function regularization eq.~(\ref{regspectr}) with 
$\lambda = 500 \ldots 800$ MeV.  The strongest effects of the cut--off are 
observed at intermediate and short distances, where the unphysical attraction 
in dimensionally regularized TPE is greatly reduced. On the other hand, 
the asymptotic behavior of the potential at large $r$ is not affected by the
choice of regularization. The results for D-- and F--waves are greatly improved 
when the spectral function regularization is used instead of dimensional one,
as documented in Fig.~\ref{fig4}. The error of about 10 (1)$^\circ$ at 
$E_{\rm lab} = 300$ MeV for the D-- (F--) waves appears reasonable.  

In ref.~\cite{EGMnew} we have demonstrated that the spectral function 
regularization is equivalent to (finite) momentum cut--off regularization 
of pion loop integrals. 
It should be understood that this new regularization scheme does 
not introduce any model dependence in the EFT procedure as soon as 
$\lambda$ is chosen of the order of (or larger than) $M_\rho$. 
Various choices for $\lambda$ (including $\lambda = \infty$, which is equivalent
to dimensional regularization) differ from each other by higher--order contact
terms and lead to exactly the same result for observables provided one 
keeps terms in all orders in the EFT expansion.
In ref.~\cite{EGMnew} we have also demonstrated how to systematically 
perform renormalization in this scheme, see also \cite{don98,don98_2}
for the related discussion. 
\begin{figure}[tb]
\begin{minipage}{5cm}
Figure 5.  The (normalized) integrand $I (\mu)$  in Eq.~(\ref{four}) for
different distances $r$.
\end{minipage}
\hskip 1.0 true cm
\begin{minipage}{6cm}
\psfig{file=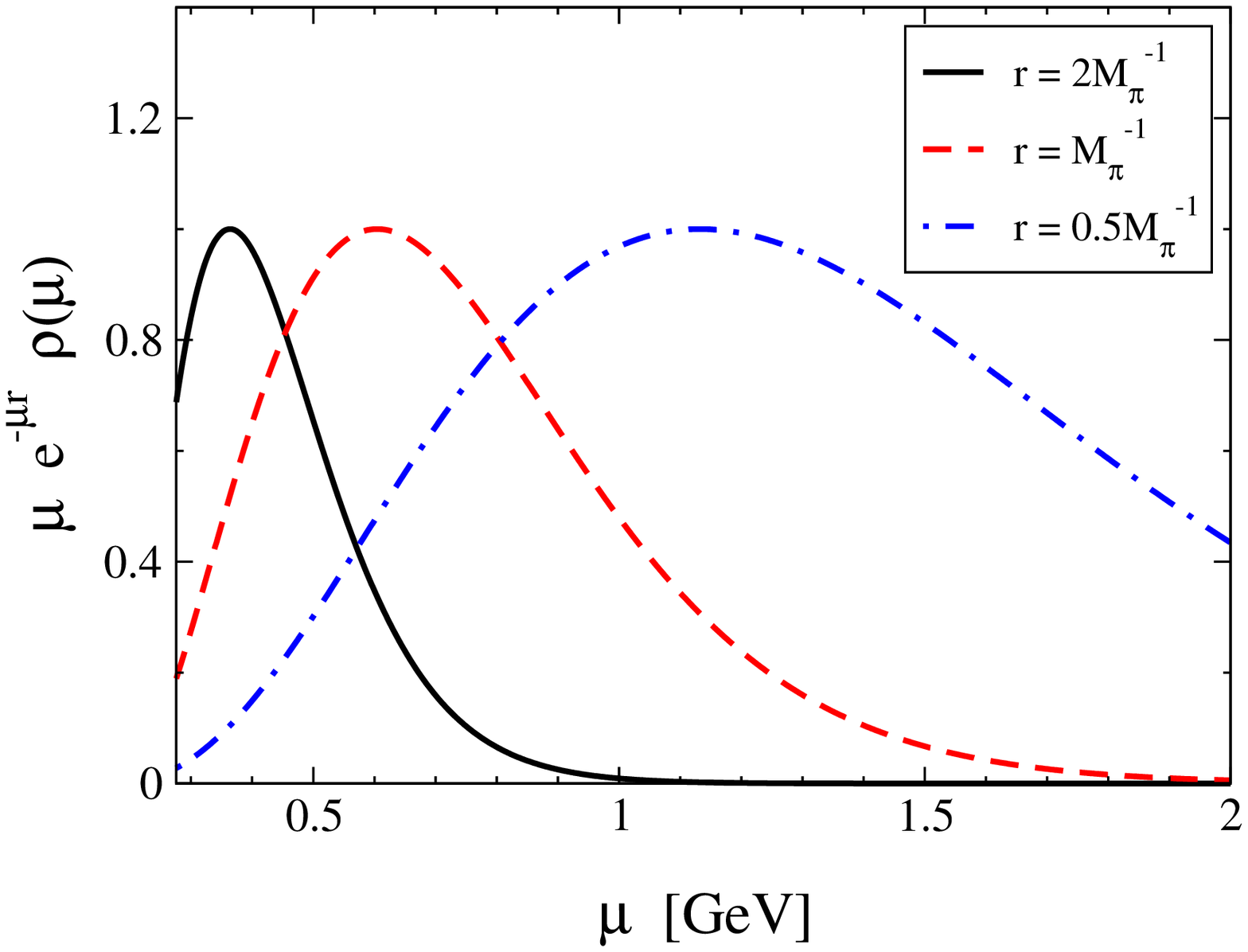,width=6cm}
\end{minipage}
\label{fig5}
\end{figure}
\setcounter{figure}{5}

Last but not least, our choice of regularization in eq.~(\ref{regspectr})
is by no means unique. Different choices lead to equivalent results for the 
potential up to higher order terms and may be used as well. The advantage 
of the form  eq.~(\ref{regspectr}) is that it does not generate spurious 
long--range contributions which are suppressed by inverse powers of $\lambda$
\cite{EGMnew}. Simple analytical expressions for the 
regularized loop functions at NNLO are given in ref.~\cite{EGMnew}.

\section{TWO NUCLEONS WITH THE NEW REGULARIZATION SCHEME}

Application of the new regularization scheme to the 
low partial waves in the non--perturbative regime is the last 
topic I would like to address. The explicit expressions for the potential up to NNLO 
derived with the new regularization scheme are given in ref.~\cite{EGMnew2}.
In that work the central values from the $Q^3$--analysis of the $\pi N$ 
system \cite{paul00} have been adopted for the LECs $c_{1,4}$: 
$c_1 = -0.81$ GeV$^{-1}$, $c_4 = 3.40$ GeV$^{-1}$. For the constant 
$c_3$ the value $c_3 = -3.40$ GeV$^{-1}$ has been used which is 
on the lower side but still consistent with the results from ref.~\cite{paul00}
($c_3 = -4.69 \pm 1.34$ GeV$^{-1}$). Interestingly, similar values for this 
LEC have been extracted recently from matching the chiral expansion 
of the nucleon mass to lattice gauge theory results at pion masses 
between 500 and 800 MeV \cite{bern03}. 

Using the effective potential the bound and scattering states are generated 
by solving the Lippmann--Schwinger (LS) equation 
which has to be regularized because of the incorrect behavior  
of the potential at large momenta (even after removing the large--mass 
components in the spectrum). We regularize the LS equation in a 
standard way using the exponential regulator function
$f^\Lambda (p) = \exp [ -p^6 / \Lambda^6]$, see \cite{EGMnew2}
for more details. In that reference $\Lambda$ has been varied 
in the range 450$\ldots$600 MeV at NLO and 450$\ldots$650 MeV
at NNLO. The cut--off in the spectral function has been varied 
independently in the range 500$\ldots$700 MeV. 
\begin{figure}[tb]
\centerline{
\psfig{file=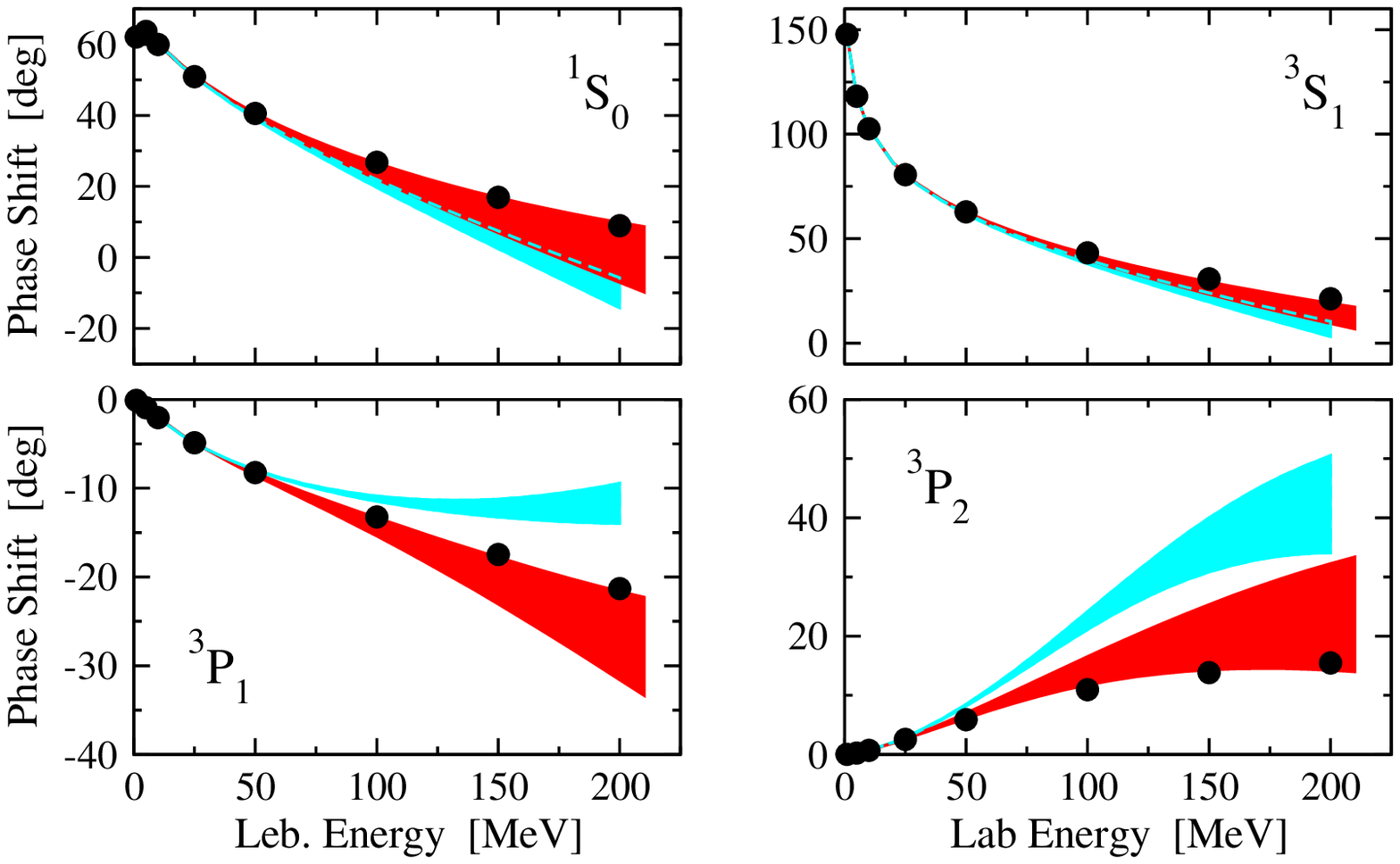,width=12.0cm}}
\vskip -0.7 true cm
\centerline{
\parbox{1.00\textwidth}{
\caption[fig6]{
\label{fig6}  
Selected low NN partial waves calculated with the new regularization scheme.
For notations see Fig.~\ref{fig4}.
}}}
\end{figure}

For any choice of $\Lambda$ and $\lambda$ (denoted in ref.\cite{EGMnew2}
by $\tilde \Lambda$) the 9 LECs accompanying the contact interactions have 
been fixed from a fit to {\it np} low partial waves at low energy 
($E_{\rm lab} < 100$ MeV) in the same way as in 
\cite{egm00,ep02}. This allows to make predictions for higher energies/partial waves. 
Results for selected partial waves 
are shown in Fig.~\ref{fig6}. The description of the data improves 
at NNLO compared to NLO in nearly all channels, which is entirely due to 
inclusion of the subleading TPE contribution. The improvement is also observed
for  the scattering lengths and effective range coefficients 
in both S--waves
as well as for the deuteron properties \cite{EGMnew2}. 
Notice that contrary to the results of 
ref.~\cite{egm00}, where DR has been applied to 
calculate pion loop integrals, there are now no spurious deeply bound states.

\section{SUMMARY AND OUTLOOK}

I have discussed several topics related to the EFT description of few--nucleon 
systems. At NNLO, the complete analysis of 2N, 3N and 4N systems 
has been performed including for the first time the chiral 3N force. 
Promising results have been obtained at NNLO for various 
$N d$ scattering observables as well as for the $\alpha$--particle binding energy. 

The problem with the unphysically strong attraction of the chiral TPE 
has been resolved using the new regularization scheme based on the 
spectral function representation. The method has already been successfully applied
to the 2N system at NNLO, where it allows to significantly improve the convergence
of the EFT expansion. It will be interesting to reconsider the 3N and 4N systems 
using this formalism.

Finally, it is now of utmost importance to investigate the next order (NNNLO) 
to be able to make conclusions about the convergence of the chiral EFT for 
nuclear systems. Work along these lines is in preparation \cite{EGMnew3}
(for a first attempt using dimensional regularization see \cite{machl03}).

\section{ACKNOWLEDGMENTS}

It is a pleasure to thank my collaborators on these topics, and particularly 
Walter Gl\"ockle, Hiroyuki Kamada, Ulf--G. Mei\3ner, Andreas Nogga and  
Henryk Wita{\l}a.

\end{document}